\documentclass[12pt]{amsart}
\usepackage{amssymb,latexsym}
\usepackage[dvips,draft]{graphicx}

\begin{document}

\title[On a time-dependent extra spatial dimension]
{On a time-dependent extra spatial dimension}
\author{Peter K.F. Kuhfittig}
\address{Department of Mathematics\\
Milwaukee School of Engineering\\
Milwaukee, Wisconsin 53202-3109, USA}
\date{\today}

\begin{abstract} In the usual brane-world scenario matter fields are confined
to the four-dimensional spacetime, called a 3-brane, embedded
in a higher-dimensional space, usually referred to as the bulk
spacetime.  In this paper we assume that the 3-brane is a de
Sitter space; there is only one extra spatial dimension,
assumed to be time dependent.  By using the form of the
brane-world energy-momentum tensor suggested by Shiromizu
\emph{et al.} in the five-dimensional Einstein
equations, it is proposed that the cosmological expansion of the
3-brane may provide a possible explanation for the collapse of
the extra dimension, as well as for the energy stored in the
resulting curled-up dimension.  More precisely, whenever the
bulk cosmological constant $\Lambda$ is negative, the extra
spatial dimension rapidly shrinks during the inflation of the brane.
When $\Lambda$ is positive, on the other hand, the extra spatial
dimension either completely follows the cosmological expansion of
the brane or completely ignores it, thereby shrinking relative
to the expanding space.  This behavior resembles the
all-or-nothing behavior of ordinary systems in an expanding
universe, as recently demonstrated by R.H. Price.
\end{abstract}

\maketitle
PACS number(s): 04.20.Cv, 04.50.+h
\vspace{12pt}

\section{Introduction}
 The notion that our world may contain more than three spatial
dimensions can be traced to the pioneer work of Kaluza and Klein
starting in 1919.  The development of string/M-theory has
resulted in a revival of this idea.  At this stage in the evolution
of the Universe the extra spatial dimensions are hidden from us
four-dimensional observers.  It has been conjectured that the extra
dimensions had suddenly compactified to become unobservable, but
the mechanisms for this dimension breaking has remained somewhat
of a mystery \cite{mK99}.  It is proposed in this paper that
cosmic inflation may provide a possible explanation, provided that
certain conditions are met.  It is also proposed that the same
mechanism would be the source of the energy stored in the
resulting curled-up dimensions.

We are going to confine ourselves to a single extra spatial
dimension with a scale factor that is necessarily time dependent
to allow the size to vary.  Accordingly, our starting point is
the spacetime topology $M\times S^1$, where $M$ refers to a de
Sitter space and $S^1$ to an extra-dimensional 1-sphere.  So our
metric is given by
\begin{equation}\label{E:line}
  ds^2=-dt^2 +[R(t)]^2\left[dr^2 +r^2(d\theta^2+\text{sin}^2
     \theta\,d\phi^2)\right]+[\rho(t)]^2 d\chi^2,
\end{equation}
where $\chi$ is the coordinate in the fifth dimension.  (Note the
time-dependent scale factor $\rho(t)$.)  In the usual
brane-world picture matter fields are confined to the
four-dimensional spacetime (or 3-brane), while gravity acts in
five dimensions (the bulk).  Our \emph{basic assumption} is that
when dealing with the Early Universe
we may apply the five-dimensional Einstein field equations
$G_{\mu\nu}=k^2_5\,T_{\mu\nu}$ using a particular form of the
energy-momentum tensor: following Ref.~\cite{SMS00},
$(M, q_{\mu\nu})$ denotes our 3-brane
in a five-dimensional spacetime $(V, g_{\mu\nu})$ and
\begin{equation}\label{E:T}
    T_{\mu\nu}=-\Lambda g_{\mu\nu}+\delta(\chi)(-\lambda q_{\mu\nu}
    +\tau_{\mu\nu}).
\end{equation}
Here $\Lambda$ is the cosmological constant of the bulk spacetime,
while the hypersurface $\chi=0$ corresponds to our 3-brane.  The
$\delta$-function expresses the confinement of matter in the brane;
its appearance here is due to the boundary surface $\chi=0,$ where
$T_{\mu\nu}$ is proportional to the $\delta$-function.  Also,
$\lambda$ and $\tau_{\mu\nu}$ are the vacuum energy and the
energy-momentum tensor, respectively, of the 3-brane.

Concerning our basic assumption, it must be kept
in mind that the coefficients in Eq.~(\ref{E:line}) are independent
of $\chi$, an assumption more in line with the Kaluza-Klein model
than the brane-world model.  It is not unreasonable to assume,
however, that during inflation the enormous rate of expansion is
so dominant that outside influences, including the existence of an
extra  spatial dimension, are negligible.  Moreover, the resulting
simplification leads directly to several interesting conclusions
that may not be apparent otherwise.

Returning to Eq.~(\ref{E:line}), if $(M,q_{\mu\nu})$ is to be a de
Sitter space, we need to let $R(t)=e^{Ht}$, where $H=
\sqrt{\Lambda_4/3}$ and $\Lambda_4$ is the cosmological constant of
the 3-brane.  Universes with exponential expansion are usually
called \emph{inflationary}.

Since the derivations in Ref.~\cite{SMS00} do not depend on the
sign of $\Lambda$, we are justified in considering the cases
$\Lambda>0$ and $\Lambda<0$ separately.  In the latter case, discussed
in Sec. \ref{S:neg}, the extra spatial dimension rapidly shrinks
during the inflation of the brane.  The former case, discussed next,
is the more interesting of the two: the extra spatial dimension either
completely follows the cosmological expansion of the brane or
completely ignores it, thereby shrinking relative to the expanding
space.

\section{The case $\Lambda>0$}\label{S:pos}

Our first step is to calculate the nonzero components of the Einstein
tensor in the orthonormal frame.  These are given next:
\begin{equation}\label{E:Einstein1}
   G_{\hat{t}\hat{t}}=3\frac{[R'(t)]^2}{[R(t)]^2}
     +3\frac{R'(t)\rho'(t)}{R(t)\rho(t)},
\end{equation}
\begin{equation}\label{E:Einstein2}
   G_{\hat{r}\hat{r}}=-2\frac{R''(t)}{R(t)}
      -\frac{[R'(t)]^2}{[R(t)]^2}-2\frac{R'(t)\rho'(t)}{R(t)\rho(t)}
      -\frac{\rho''(t)}{\rho(t)},
\end{equation}
\begin{equation}\label{E:Einstein3}
   G_{\hat{\theta}\hat{\theta}}=G_{\hat{\phi}\hat{\phi}}
   =-2\frac{R''(t)}{R(t)}-\frac{[R'(t)]^2}{[R(t)]^2}
    -\frac{\rho''(t)}{\rho(t)}-2\frac{R'(t)\rho'(t)}{R(t)\rho(t)},
\end{equation}
\begin{equation}\label{E:Einstein4}
  G_{\hat{\chi}\hat{\chi}}
    =-3\frac{R''(t)}{R(t)}-3\frac{[R'(t)]^2}{[R(t)]^2}.
  \end{equation}
(The components of the Riemann curvature and Ricci tensors are given
in the Appendix.)

\subsection{Solutions}

From Eq.~(\ref{E:Einstein1}), we have
\begin{equation*}
  3\frac{[He^{Ht}]^2}{[e^{Ht}]^2}+3\frac{He^{Ht}\rho'(t)}
    {e^{Ht}\rho(t)}=k^2_5\,T_{\hat{t}\hat{t}},
\end{equation*}
which reduces to
\begin{equation}\label{E:equation1}
  \frac{\rho'(t)}{\rho(t)}=-H+\frac{k^2_5\,T_{\hat{t}\hat{t}}}
    {3H}.
\end{equation}
Let $A_{\text{in}}$ be the initial value of $\rho(t)$ (at the
onset of inflation), i.e., $\rho(0)=A_{\text{in}}$.  Then the
solution is
\[
    \rho(t)=A_{\text{in}}e^{-Ht}e^{k^2_5\,T_{\hat{t}\hat{t}}t/3H}.
\]
By Eq.~(\ref{E:T}), since $\delta(\chi)=0$ in the bulk,
\begin{equation}\label{E:solution1}
   \rho(t)=A_{\text{in}}e^{-Ht}e^{k^2_5\,\Lambda t/3H}.
\end{equation}

Similarly, from both Eqs.~(\ref{E:Einstein2}) and (\ref{E:Einstein3}),
we get
\begin{equation}\label{E:equation2}
  \rho''(t)+2H\rho'(t)+(3H^2-k^2_5\,\Lambda)\rho(t)=0
\end{equation}
and
\begin{equation}\label{E:solutiongeneral}
  \rho(t)=c_1\,e^{-Ht}e^{\sqrt{-2H^2+k^2_5\,\Lambda}\,t}
   +c_2\,e^{-Ht}e^{-\sqrt{-2H^2+k^2_5\,\Lambda}\,t},
\end{equation}
where $c_1$ and $c_2$ are arbitrary constants. Solutions
(\ref{E:solution1}) and (\ref{E:solutiongeneral}) must agree,
particularly at $t=T$, the end of inflation.  It follows that
$c_1=A_{\text{in}}$ and $c_2=0$.  The second term can also be
eliminated for physical reasons: since $HT\approx 100$
\cite{tR93, KT90}, $e^{-HT}e^{-\sqrt{-2H^2+k^2_5\,\Lambda}\,T}$
is many orders of magnitude below Planck length and would
therefore not make physical sense.  So
\begin{equation}\label{E:solution2}
  \rho(t)=A_{\text{in}}e^{-Ht}
    e^{\sqrt{-2H^2+k^2_5\,\Lambda}\,t}.
   \end{equation}
Eqs.~(\ref{E:solution1}) and (\ref{E:solution2}) now yield
\begin{equation}
    \frac{k^2_5\,\Lambda}{3H}=\sqrt{-2H^2+k^2_5\,\Lambda}
\end{equation}
with $H=\sqrt{\Lambda_4/3}$.  The only solutions are $\Lambda=
\Lambda_4/k^2_5$ and $\Lambda=\Lambda_4/(\frac{1}{2}k^2_5)$.

It remains to show that these solutions are consistent with
$G_{\hat{\chi}\hat{\chi}}$.  Observe that this component is
completely independent of $\rho(t)$, suggesting that $\chi=0$ in
Eq.~(\ref{E:T}). Retaining the notation $\delta(\chi)$,
Eq.~(\ref{E:T}) becomes
\begin{equation}\label{E:T5}
  T_{\hat{\chi}\hat{\chi}}=-\Lambda g_{\hat{\chi}\hat{\chi}}
    +\delta(\chi)(-\lambda q_{\hat{\chi}\hat{\chi}}
         +\tau_{\hat{\chi}\hat{\chi}}),
\end{equation}
and, since $R(t)=e^{Ht},$ with $H=\sqrt{\Lambda_4/3}$,
\begin{equation}
     G_{\hat{\chi}\hat{\chi}}=-3\frac{R''(t)}{R(t)}
      -3\frac{[R'(t)]^2}{[R(t)]^2}=-\Lambda_4-\Lambda_4.
\end{equation}
In Eq.~(\ref{E:T5}),
$q_{\hat{\chi}\hat{\chi}}=0$ and
$g_{\hat{\chi}\hat{\chi}}=1$, so that
\begin{equation}\label{E:T6}
   -\Lambda_4-\Lambda_4=k^2_5[-\Lambda+\delta(\chi)
     \tau_{\hat{\chi}\hat{\chi}}].
\end{equation}
It now follows that $\delta(\chi)\tau_{\hat{\chi}\hat{\chi}}
=0$ if, and only if, $\Lambda=\Lambda_4/(\frac{1}{2}k^2_5)$.
So if $\delta(\chi)\tau_{\hat{\chi}\hat{\chi}}\ne 0$, then
$\Lambda=\Lambda_4/k^2_5$, the other solution.  For this case,
Eq.~(\ref{E:T6}) implies that
\begin{equation}\label{E:othersolution}
   -2\Lambda_4=k_5^2[-\Lambda_4/k_5^2+\delta(\chi)
      \tau_{\hat{\chi}\hat{\chi}}].
\end{equation}
One of the properties of the $\delta$-function is that
$\delta(y)f(y)=\delta(y)f(0)$ for any continuous function $f$.
Applying this property to $\tau_{\hat{\chi}\hat{\chi}}$, we have
\begin{equation*}
    \delta(\chi)\tau_{\hat{\chi}\hat{\chi}}=
    \delta(\chi)\left(\tau_{\hat{\chi}\hat{\chi}}\left|
     _{\chi=0}\right)\right.,\phantom{e^5}
\end{equation*}
emphasizing the confinement to the brane.  So
Eq.~(\ref{E:othersolution}) becomes
\[
   \delta(\chi)\left(\tau_{\hat{\chi}\hat{\chi}}\left|
     _{\chi=0}\right)\right.=-\frac
      {\Lambda_4}{k_5^2}.\phantom{e^5}
\]
This form is similar to the classical de Sitter forms
\[
    T_{\hat{r}\hat{r}}=T_{\hat{\theta}\hat{\theta}}=
     T_{\hat{\phi}\hat{\phi}}=
          -\frac{\Lambda_4}{8\pi}
\]
obtainable from Eqs.~(\ref{E:Einstein2}) and (\ref{E:Einstein3})
by letting  $k_5^2=8\pi$ and eliminating $\rho(t)$.

\subsection{Analysis}
We now examine each solution in turn:

\textbf{(1)} Suppose $\Lambda=\Lambda_4/k^2_5$.  Substituting in Eqs.~
(\ref{E:solution1}) and (\ref{E:solution2}) and recalling
that $H=\sqrt{\Lambda_4/3}$, we get in both cases,
\begin{equation}\label{E:fixed}
   \rho(t)=A_{\text{in}}e^{-Ht}e^{Ht}=A_{\text{in}}.
\end{equation}

\vspace{14pt}
\textbf{(2)} For the other solution, $\Lambda=\Lambda_4/(\frac{1}{2}k^2_5)$,
we obtain from Eqs.~(\ref{E:solution1}) and (\ref{E:solution2}),
\begin{equation}
  \rho(t)=A_{\text{in}}e^{-Ht}e^{2Ht}=A_{\text{in}}e^{Ht}.
\end{equation}
So the ``radius" of the extra dimension expands by the factor
$e^{Ht}$, the same as for any other distance in the brane.
To see this, consider the proper circumference $C$ of the circle
in the ``equatorial slice" $\theta=\pi/2$ of the sphere $r=a$:
\[
   C=\int_0^{2\pi}e^{Ht}a\,d\phi=e^{Ht}(2\pi a),
          \qquad t=\text{constant}.
\]
So $C$ is simply $e^{Ht}$ times the initial proper circumference.
Similarly, the radial proper distance separating two points $A$
and $B$ for any fixed $t$ is given by
\[
  \ell(t)=\int_{r_A}^{r_B}e^{Ht}dr=e^{Ht}(r_B-r_A).
\]
So $\ell(t)$ is just $e^{Ht}$ times the initial proper radial
separation.

\vspace{14pt}
To emphasize the rescaling of the $r$ coordinate on each $t=$
constant slice, one could write
\[
   \frac{\ell(t)}{e^{Ht}}=r_B-r_A.
\]
In the first case, Eq.~(\ref{E:fixed}), the corresponding
equation for a fixed $\ell_1(t)=r_B-r_A$ can be written
\[
   \frac{\ell_1(t)}{e^{Ht}}=e^{-Ht}(r_B-r_A),
\]
that is, if $\rho(t)$ remains fixed, then the original
size has the appearance of having shrunk by a factor of $e^{-HT}$
relative to the expanded space at the end of inflation.

The shrinking case does have one obvious consequence: for the
extra dimension to be very small,
say Planck size, at the end of inflation, we must have
\[
  A_{\text{in}}e^{-HT}\approx A_{\text{in}}e^{-100}\approx
    10^{-35}\,\text{m},
\]
leading to the rather large value $A_{\text{in}}\approx
2.7\times10^8\,\text{m}$. Not being part of the 3-brane,
such a large extra dimension cannot be ruled out.
Consider also the following model proposed by Chodos and
Detweiler (1980 \cite{CD80}):
\begin{equation}
  ds^2=-dt^2+\left(\frac{t}{t_0}\right)(dx^2+dy^2+dz^2)
         +\frac{t_0}{t}d\chi^2.
\end{equation}
Here the size of the extra spatial dimension exceeds that of the
other three for $t<t_0$.  In fact, at the singularity $t=0$, there
is only one spatial dimension whose size approaches infinity.

The last possibility is actually an exception to the shrinking case:
if $A_{\text{in}}$ is infinitely large to start with, it remains
infinitely large.  This outcome may be more than just a curiosity:
an infinite extra dimension is required in certain models
\cite{RS2}.

\section{The case $\Lambda<0$}\label{S:neg}
If $\Lambda<0$, the solution to Eq.~(\ref{E:equation1})
retains the form
\begin{equation}\label{E:solution3}
  \rho(t)=A_{\text{in}}e^{-Ht}e^{k^2_5\,\Lambda t/3H}.
\end{equation}
As noted earlier, at the end of inflation $(t=T)$, we have
$HT\approx 100$, so that $\rho(t)$ is now
many orders of magnitude below Planck length and is therefore
not an acceptable solution.  But from Eq.~(\ref{E:equation2})
we get
\begin{equation}\label{E:solution4}
   \rho(t)=e^{-Ht}\left(A_{\text{in}}\,\,
    \text{cos}{\sqrt{2H^2-k^2_5\,\Lambda}\,t}\right.
   \left.+\,c_2\,\,\text{sin}{\sqrt{2H^2-k^2_5\,\Lambda}\,t}\right).
\end{equation}
This solution is plausible, based on the discussion at the end of
Sect.~\ref{S:pos}, but it also shows that $\rho(t)$ has shrunk
significantly at the end of inflation.

\section{The stored energy}

It is generally believed that the extra spatial dimensions
are tightly curled up, thereby storing huge amounts of
potential energy.  This stored energy may actually be the
source of \emph{dark energy} that causes the expansion
of the Universe to accelerate.  The existence of such
energy is confirmed in the present model.  Consider, for
example, the radial tidal constraint
$|R_{\hat{t}\hat{r}\hat{r}}^{\phantom{\hat{t}\hat{t}
\hat{r}}\hat{t}}|$ \cite{MT88}.  From the Appendix,
\[
  |R_{\hat{t}\hat{r}\hat{r}}
    ^{\phantom{\hat{t}\hat{t}\hat{r}}\hat{t}}|=
    \left|-\frac{R''(t)}{R(t)}\right|=H^2=
     \frac{\Lambda_4}{3}.
\]
Since $HT\approx 100$, we have $H(10^{-34}+10^{-32})
\approx 100$, so that $H\approx 10^{34}$ \cite {tR93}.
So $|R_{\hat{t}\hat{r}\hat{r}}
    ^{\phantom{\hat{t}\hat{t}\hat{r}}\hat{t}}|\approx
10^{34} /3$.  As the Universe keeps expanding,
$\Lambda_4/3$ becomes very small.  Suppose, on the other
hand, that a very tiny observer moves radially in the
extra-dimensional 1-sphere.  Since $\rho(t)=
A_{\text{in}}e^{-Ht}$, the tidal constraint (from the
Appendix) is given by
\[
    |R_{\hat{t}\hat{\chi}\hat{\chi}}
    ^{\phantom{\hat{t}\hat{\chi}\hat{\chi}}\hat{t}}|=
    \frac{\Lambda_4}{3}=\frac{10^{34}}{3}.
\]
Unlike the expanding case discussed above, $\rho(t)$
cannot continue to shrink beyond Planck length,
thereby terminating with a huge value for the tidal
constraint.  This points to a large amount of stored
potential energy, the source of which is inflation.

\section{Summary}

It is proposed in this paper that in the case of a  de Sitter
3-brane world, cosmic inflation may provide an explanation for
the collapse of the extra spatial dimension, as well as
the source of the energy stored in the resulting curled-up
dimension.  It is shown that whenever the cosmological
constant $\Lambda$ is negative, $\rho(t)$ shrinks rapidly
during the inflation of the brane. When $\Lambda$ is positive,
the extra spatial dimension either completely follows  the
cosmological expansion of the brane or completely ignores it.
In the former case the extra dimension expands, rather than
shrinks.  In the latter case, $\rho(t)$ shrinks relative to
the expanding space, unless the size of the extra dimension
is infinite to start with.

For $\Lambda>0$, the conclusion resembles the interesting
all-or-nothing behavior demonstated in Ref.~ \cite{rP05}: a
system will either completely follow the cosmological
expansion of the universe or completely ignore it.

\section*{Appendix}

The nonzero components of the Riemann curvature tensor are
\begin{equation*}
   R_{\hat{t}\hat{r}\hat{r}}
    ^{\phantom{\hat{t}\hat{t}\hat{r}}\hat{t}}
    =R_{\hat{t}\hat{\theta}\hat{\theta}}
    ^{\phantom{\hat{t}\hat{\theta}\hat{\theta}}\hat{t}}
    =R_{\hat{t}\hat{\phi}\hat{\phi}}
    ^{\phantom{\hat{t}\hat{\phi}\hat{\phi}}\hat{t}}
     =-\frac{R''(t)}{R(t)},
\end{equation*}
\begin{equation*}
   R_{\hat{t}\hat{\chi}\hat{\chi}}
    ^{\phantom{\hat{t}\hat{\chi}\hat{\chi}}\hat{t}}
    =-\frac{\rho''(t)}{\rho(t)},
\end{equation*}
\begin{equation*}
   R_{\hat{r}\hat{\theta}\hat{\theta}}
   ^{\phantom{\hat{r}\hat{\theta}\hat{\theta}}\hat{r}}=
    R_{\hat{r}\hat{\phi}\hat{\phi}}
     ^{\phantom{\hat{r}\hat{\phi}\hat{\phi}}\hat{r}}=
    R_{\hat{\theta}\hat{\phi}\hat{\phi}}
     ^{\phantom{\hat{\theta}\hat{\phi}\hat{\phi}}\hat{\theta}}=
     -\frac{[R'(t)]^2}{[R(t)]^2},
\end{equation*}
\begin{equation*}
   R_{\hat{r}\hat{\chi}\hat{\chi}}
    ^{\phantom{\hat{r}\hat{\chi}\hat{\chi}}\hat{r}}=
     R_{\hat{\theta}\hat{\chi}\hat{\chi}}
     ^{\phantom{\hat{\theta}\hat{\chi}\hat{\chi}}\hat{\theta}}=
     R_{\hat{\phi}\hat{\chi}\hat{\chi}}
     ^{\phantom{\hat{\phi}\hat{\chi}\hat{\chi}}\hat{\phi}}=
      -\frac{R'(t)\rho'(t)}{R(t)\rho(t)}.
\end{equation*}

The components of the Ricci tensor are
\begin{equation*}
  R_{\hat{t}\hat{t}}=-3\frac{R''(t)}{R(t)}-\frac{\rho''(t)}{\rho(t)},
\end{equation*}
\begin{equation*}
  R_{\hat{r}\hat{r}}=R_{\hat{\theta}\hat{\theta}}
    =R_{\hat{\phi}\hat{\phi}}=\frac{R''(t)}{R(t)}
   +2\frac{[R'(t)]^2}{[R(t)]^2}
     +\frac{R'(t)\rho(t)}{R(t)\rho(t)},
\end{equation*}
\begin{equation*}
   R_{\hat{\chi}\hat{\chi}}=3\frac{R'(t)\rho'(t)}{R(t)\rho(t)}
     +\frac{\rho''(t)}{\rho(t)}.
\end{equation*}

\end{document}